\documentclass[%
 reprint,
superscriptaddress,
 amsmath,amssymb,
 aps,
floatfix,
]{revtex4-1}

\usepackage{graphicx}
\usepackage{dcolumn}
\usepackage{bm}
\usepackage{placeins}

\begin{document}

\preprint{APS/123-QED}

\title{Geometry, packing, and evolutionary paths to increased multicellular size}

\author{Shane Jacobeen}
 \altaffiliation{Co-first author}
 \affiliation{School of Physics, Georgia Institute of Technology, North Ave NW, Atlanta, GA 30332}
\author{Elyes C. Graba}	
 \altaffiliation{Co-first author}
 \affiliation{School of Physics, Georgia Institute of Technology, North Ave NW, Atlanta, GA 30332}
\author{Colin G. Brandys}
 \affiliation{School of Physics, Georgia Institute of Technology, North Ave NW, Atlanta, GA 30332}
\author{T. Cooper Day}
 \affiliation{School of Physics, Georgia Institute of Technology, North Ave NW, Atlanta, GA 30332}
\author{William C. Ratcliff}\affiliation{School of Biological Sciences, Georgia Institute of Technology, North Ave NW, Atlanta, GA 30332}
\author{Peter J. Yunker}
 \email{peter.yunker@physics.gatech.edu}
 \affiliation{School of Physics, Georgia Institute of Technology, North Ave NW, Atlanta, GA 30332}

\date{\today}

\begin{abstract}
The evolutionary transition to multicellularity transformed life on earth, allowing for the evolution of large, complex organisms. While multicellularity can be strongly advantageous, its earliest stages bring unique physical challenges. Nascent multicellular organisms must contend with a novel constraint: intercellular stresses arising from cell-cell interactions that can limit multicellular size. Among the possible evolutionary routes to overcoming this size limit, two appear obvious: multicellular organisms can increase intercellular bond strength, allowing them to tolerate larger stresses, or, they can slow the rate of stress accumulation by altering their internal structure. Recent experiments demonstrated that multicellular `snowflake yeast' readily find a solution to this problem via the latter route. By evolving more elongated cells, which decreases cellular packing fraction and thus the rate of internal stress accumulation during growth, snowflake yeast evolve to delay fracture and grow larger. However, it is unclear if snowflake yeast evolved large size along an optimal path, or if the observed path to large size was taken due to proximate biological reasons. Here, we examine the geometric efficiency of both strategies using a minimal model that was previously demonstrated to replicate many experimentally observed structural properties of snowflake yeast. We find that changing geometry is a far more efficient route to large size than evolving increased intercellular adhesion. In fact, increasing cellular aspect ratio is on average $\sim13$ times more effective at increasing the number of cells in a cluster than increasing bond strength. These results suggest that geometrically-imposed physical constraints may have been a key early selective force guiding the emergence of multicellular complexity.

\end{abstract}

\pacs{Valid PACS appear here}
\maketitle


\section{\label{sec:level1}Introduction}

The evolution of multicellular organisms from single-celled ancestors set the stage for unprecedented increases in complexity, especially in plants and animals \cite{12step_2005, knoll2011multiple}. In nascent multicellular organisms, size and complexity are strongly related \cite{Willensdorfer_2008, 12step_2005}; recent work has highlighted the potential for a size-complexity evolutionary feedback loop \cite{knoll2011multiple}. However, it is unclear how early, simple multicellular organisms evolved to be larger. Newly multicellular organisms lack genetically-regulated development, growing instead through the stochastic replication of physically-attached individual cells. At high cell densities, stochastic growth can result in large intercellular forces \cite{Hallatschek2016}, fragmenting groups and limiting multicellular size \cite{Jacobeen_2017}. Thus, mitigating internal mechanical stress is one of the first evolutionary challenges faced by nascent multicellular organisms. Though the transition to multicellularity occurred independently in at least 25 separate lineages \cite{Bonner_1998,GrosbergStrathmann2007}, we know little about the physical properties of early multicellular lineages due to their ancient origins and limitations of the fossil record. 

Nonetheless, there are two clear routes to increased size in nascent multicellular clusters of cells whose size is limited by the accumulation of internal stress: an organism could evolve to withstand larger intercellular stresses, or, it could evolve to accumulate intercellular stresses at a slower rate during growth. The former strategy would likely involve evolving stronger intercellular bonds, while the later would involve changes to structural geometry. Geometrically-imposed physical constraints play key roles in the organization of numerous microbial systems, including growing biofilms and swarming or swimming communities \cite{Smith2016Cell, ilkanaiv_effect_2017, guadayol_cell_2017, PhysRevLett.119.188101}. Separating geometric effects from biological processes is nontrivial \cite{Seppe_Pareto}, however, and little is known about how simple multicellular systems respond to selection for increased size.

Recently, model systems of simple multicellularity have allowed the early steps of this transition to be studied in the lab with unprecedented precision \cite{Ratcliff2012Experimental, HERRONetal2013, KOSCHWANEZetal2013, ratcliff2013experimental}. In the case of `snowflake yeast' \cite{Ratcliff2012Experimental, Ratcliff2012Experimental}, simple multicellular clusters of \textit{Saccharomyces cerevisiae} are subjected to daily selection for large size; they rapidly evolve to double their maximum number of cells per cluster in just seven weeks \cite{Jacobeen_2017}. Snowflake yeast cluster size is limited by the fracturing of intercellular bonds under growth-induced stresses (Figure 1a). Larger size at fracture is accomplished primarily by a simple change to cluster geometry: over $\sim291$ generations, snowflake yeast evolved to have more elongated cells. This increase in cellular aspect ratio decreases the cellular packing fraction, slowing the accumulation of internal stress and delaying fracture \cite{Jacobeen_2017} (Figure 1b). Cellular elongation is a highly parallel evolutionary trait, evolving independently in replicate populations \cite{ratcliff_tempo_2013}. However, it remains unclear \textit{why} this evolutionary route to large size is repeatedly observed: do snowflake yeast clusters modify geometry because it is more effective than increasing the strength of cell-cell bonds, or for proximate reasons relating to the model system (e.g., perhaps there are more possible mutations affecting geometry). 

To investigate the roles of geometry and bond strength in the evolution of nascent multicellularity, we employ a geometric model of experimentally-evolved snowflake yeast \cite{Ratcliff2012Experimental, ratcliff_tempo_2013, Ratcliff2015Origins}, introduced and experimentally validated in Jacobeen \textit{et. al.}, 2017 \cite{Jacobeen_2017}. We find that modifying packing geometry, and thus slowing the accumulation of internal stresses, is a far more efficient route to large size than increasing intercellular bond strength. This result is likely general, as cells are capable of imparting tremendous forces during growth \cite{delarue2016self}, and the resulting cell-cell forces increase rapidly in jammed aggregates. Thus, evolving physical robustness by modifying multicellular geometry may have been a key early selective force guiding the emergence of multicellular complexity.

\begin{figure}[!tbh]
\centering
\includegraphics[width=\linewidth]{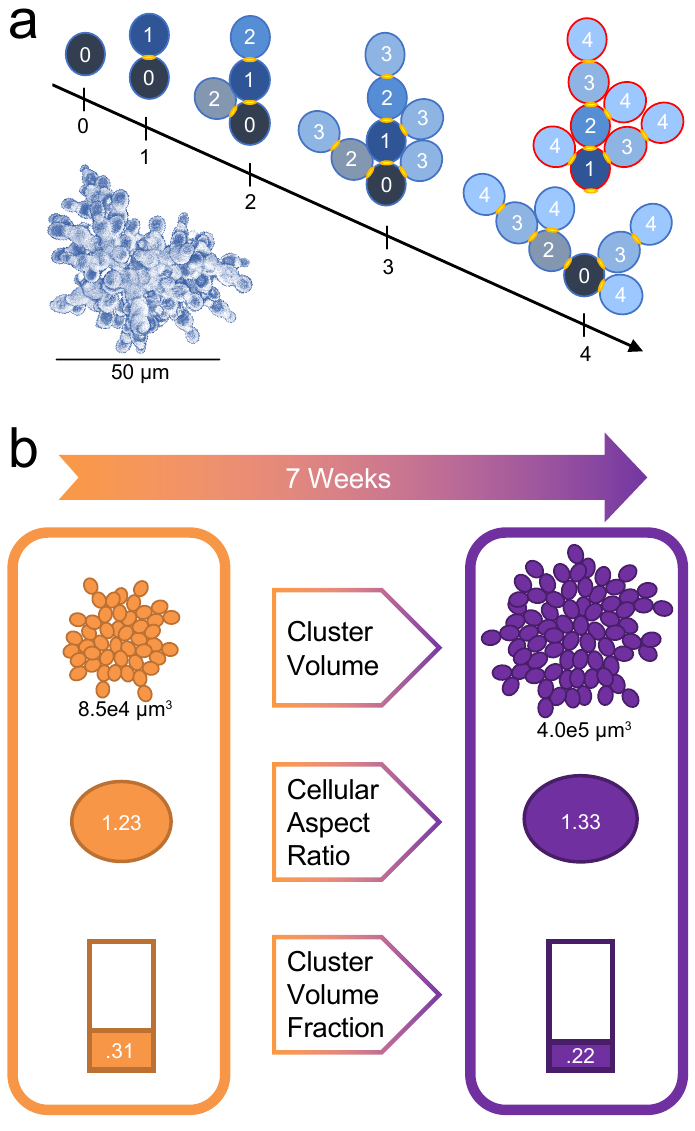}
\caption{(a) 2D schematic of snowflake yeast growth morphology, showing fracture due to cellular crowding. Inset: 3D confocal image of a snowflake yeast cluster. (b) Changes over 7 weeks of experimental evolution in mean values of snowflake yeast cluster size, cellular aspect ratio, and cluster volume fraction.}
\end{figure}

\section{Background}

We simulate the growth of snowflake yeast clusters with a simple, three-dimensional geometric model \cite{Jacobeen_2017} based on their fractal-like growth pattern \cite{Ratcliff2012Experimental}. The model is purely structural, \textit{i.e.}, it lacks dynamics, yet it accurately reproduces many relevant experimentally-measured structural properties of snowflake yeast \cite{Jacobeen_2017} (for more details on experimental validation of the model, please see the Supplemental Information).
    
\subsection{\label{sec:level2}Model}

Snowflake yeast cells reproduce via budding \cite{Ratcliff2012Experimental}; daughter cells remain attached to their mothers, creating a biologically and physically tractable multicellular cluster (Fig. 1a). In our simulation, cells are modeled as prolate spheroids (ellipsoids in which two `equatorial' radii are equal and less than the polar radius), with major-minor axis aspect ratio $\alpha$.  Each generation, all cells in the cluster attempt to reproduce by adding a daughter cell of identical volume on their surface. Daughter cells are placed at a specified angle from the polar axis, called the angle of attachment, $\theta$, where $\theta$ is the acute angle between the parent cell's major axis and a vector that originates at the geometric center of the cell and passes through the point on its surface at which the daughter cell attaches. Thus, daughter cells are randomly placed  along a `budding ring' on their parent's surface. Additionally, cells other than the basal cell have an $80\%$ chance of spawning at the pole opposite their parent (that is, with $\theta = 0$) on their initial reproduction attempt. Cellular bodies may overlap, but the center-to-center separation may not be less than $50\%$ of their small diameter; this maximum strain condition is analogous to disallowing the overlap of bud scars (\textit{i.e.}, attachment sites). If the randomly selected attachment site would cause too much ($>$$50\%$) overlap, the daughter cell is not created and the parent cell misses their chance to reproduce that generation.

Varying $\theta$ and $\alpha$ facilitate changes to cluster geometry. To vary bond strength, we first calculate the deformation energy ($u$) between the bodies of neighboring cells. That is
\begin{equation}
u_{ij} = (d - r_{i} - r_{j})^2
\end{equation}
where $d$ is the center-to-center distance between overlapping cells, and $r_i$ and $r_j$ are the equatorial radii of two neighboring cells. $u_{ij} = 0$ for non-overlapping cells, and the total `deformation energy' ($U$) in a cluster is the sum of individual $u_{ij}$:
\begin{equation}
U = \sum_{i=1}^{N}\sum_{j\neq i}^{N} u_{ij}
\end{equation}
where N is the number of cells in the cluster. In a real cluster, cells would bend at their cell-cell bonds rather than overlap, so linear overlap acts as a proxy for deformation and squared overlap is a proxy for deformation energy, or internal stress within the cluster (using a Hertzian, rather than a harmonic model for deformation energy does not qualitatively change the results of this simulation \cite{Jacobeen_2017}). As clusters fracture due to an accumulation of internal stress \cite{Jacobeen_2017}, we use a $U$ threshold ($U_{c}$) to limit cluster size. Snowflake clusters fracture when their internal stress exceeds the ultimate strength of the cell-cell bonds; thus, changing $U_{c}$ is analogous to changing bond strength.

\begin{table}[!tbh]
  \centering
\begin{tabular}{ |l|l| }
  \hline
  \multicolumn{2}{|c|}{Abbreviation Directory} \\
  \hline
  $\alpha$ & Cellular Aspect Ratio \\
  $\theta$ & Angle of attachment \\
  $U$ & Deformation Energy \\
  $U_{c}$ & Deformation Energy Threshold (ultimate bond strength) \\
  $N$ & Number of cells in a cluster \\
  \hline
\end{tabular}
  \caption{A list of the various model parameters and their abbreviations} \label{tab:sometab}
\end{table}

As previously reported in Jacobeen \textit{et. al.}, 2017 \cite{Jacobeen_2017}, this geometric model recapitulates many key structural features observed in experiments. Experimentally evolved isolates were modeled by randomly picking each new cell's $\alpha$ from experimentally measured distributions. These simulations revealed that as mean cellular $\alpha$ increases, cluster volume fraction decreases. In fact, simulations closely replicate experimental observations: simulated and experimentally measured packing fractions are within 5\% of each other for all four strains studied (the validation of the model via comparison with experimental results is detailed in \cite{Jacobeen_2017} and in the Supplemental Information here). As internal stress limits cluster size by fracturing intercellular bonds, the decrease in volume fraction due to cellular geometry modification likely plays a large role in the evolved increase in cluster volume over seven weeks \cite{Jacobeen_2017} (Figure 1b).

\section{Results}

\begin{figure*}[!tbh]
\centering
\includegraphics[width=\textwidth]{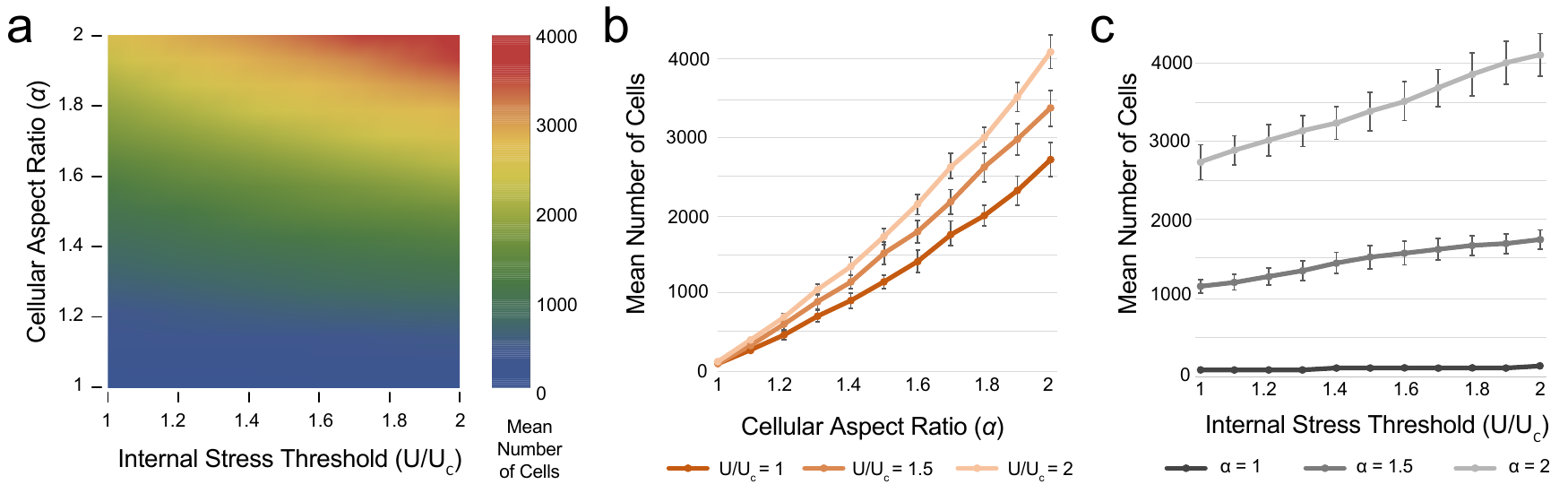}
\caption{(a) Interpolated heat map of the mean number of cells in a cluster as a function of cellular aspect ratio ($\alpha$) and deformation energy threshold ($U_{c}$). (b) Mean number of cells per cluster versus $\alpha$ for $U_{c}$ (dark orange), $1.5U_{c}$ (medium orange), and $2U_{c}$ (light orange). (c) Mean number of cells per cluster versus $U_{c}$ for $\alpha=1.0$ (dark gray), $\alpha=1.5$ (medium gray), and $\alpha=2.0$ (light gray).  Each data point is the average of 100 independent simulations. Error bars indicate standard deviation.}
\end{figure*}
\paragraph*{}

\begin{figure}[!tbh]
\centering
\includegraphics[width=\linewidth]{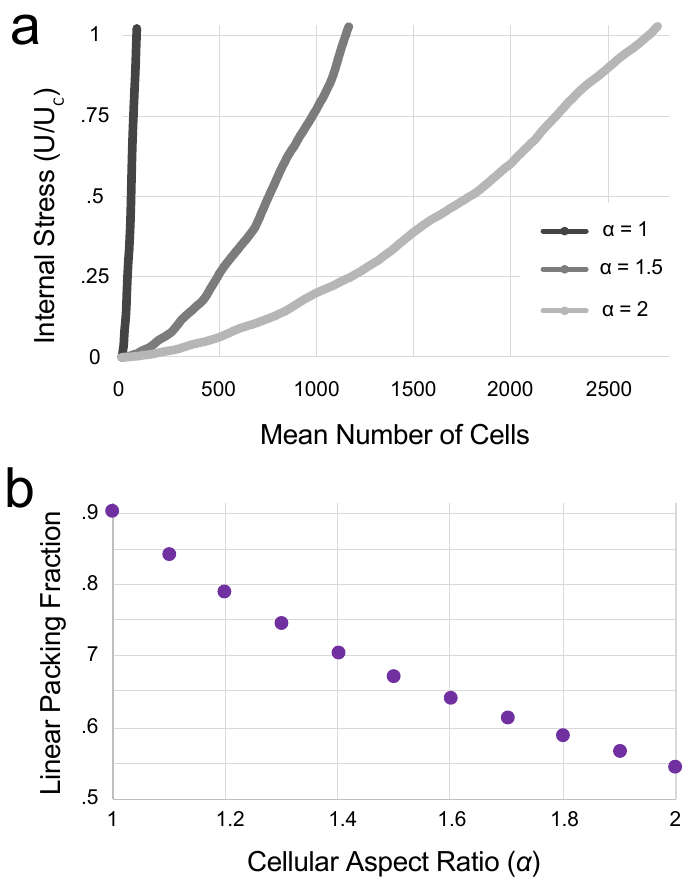}
\caption{(a) As a cluster grows, total deformation energy, $U$,increases as well. This increase is rapid when $\alpha = 1$ (dark gray), moderate for $\alpha = 1.5$ (medium gray), and slowest for $\alpha = 2 $ (light gray). Each overlappingn data point is the average of 100 independent simulations. (b) Linear packing fraction for $5$ daughter cells on a single mother cell as a function of aspect ratio $\alpha$ ($\theta=54^{\circ}$).}
\end{figure}

To directly compare the efficiency of increasing cluster size via cellular elongation and increased bond strength, we first simulated clusters with a wide range of $\alpha$ and $U_{c}$ values (we set $\theta=54^{\circ}$, as cluster size is maximized for this value). We varied $\alpha$ and $U_{c}$ between 1.0 and 2.0 in steps of 0.1, and simulated 100 clusters for each pair of parameters (Fig. 2a). The mean number of cells per cluster increases rapidly with increasing $\alpha$ for any value of $U_{c}$ (Fig. 2a and b). In contrast, the mean number of cells increases much more slowly with increasing $U_{c}$ (Figure 2a and c). Thus increasing $\alpha$ is a more efficient path to large size than increasing $U_c$.

While increasing $\alpha$ always increases cluster size more then increasing $U_{c}$, the size of this disparity varies. For example, the smaller $\alpha$ is, the more beneficial it is to increase $\alpha$ than $U_c$. In fact, for clusters of spherical cells ($\alpha=1.0$), it is on average $\sim59$ times more effective to increase $\alpha$ than to increase $U_{c}$ (\textit{i.e.}, for small $\alpha$, there is almost no discernible gradient along the $U_{c}$ axis (Fig. 2a)). Thus, there is an especially large incentive to increase aspect ratio at least a little above $1.0$. Further, increasing $U_{c}$ always enlarges the incentive for increasing $\alpha$; this is visible in Fig. 2a as the strength of the vertical gradient increases with $U_{c}$. Though the relative superiority of increasing $\alpha$ over $U_{c}$ varies over the studied range of parameters -- generally decreasing significantly with increasing $\alpha$ and increasing with $U_{c}$ -- it is always at least $2.5$ times more effective to increase $\alpha$, and on average $\sim13$ times more cells are added for an increase of $.1$ in $\alpha$ than for an increase of the same magnitude in $U_{c}$.

Why is increasing aspect ratio a more efficient route to large size than increasing bond strength? To investigate, we measured the deformation energy in simulated clusters as a function of the mean number of cells. $U$ increases $\sim$quadratically with $N$ for any value of $\alpha$ (Figure 3a). Thus, increasing $U_c$ yields sub-linear returns ($N \sim \sqrt{U_c}$). However, increasing $\alpha$ causes $U$ to increase at a slower rate, allowing more cells to be added before $U_c$ is reached. The linear relationship between $N$ and $\alpha$ (Fig. 2c) further demonstrates the superior returns on increasing $\alpha$ rather than $U_c$.

To understand how cellular aspect ratio affects internal sress accumulation, we calculated the linear packing fraction (\textit{i.e.}, the occupied fraction of the budding ring) of 5 non-overlapping daughter cells on a parent cell for $\theta=54^{\circ}$ (5 cells was chosen because it is the maximum number that can be placed at $\theta=54^{\circ}$ for all values of $\alpha$ between 1 and 2) (Figure 3b). Larger $\alpha$ daughter cells have smaller widths; smaller widths make it less likely for any two cells to overlap. Thus, more cells must be added to clusters with large $\alpha$ to obtain the same packing fraction - and $U$ - as clusters with small $\alpha$.

We also investigated other geometric parameters, to determine if the effects of $\alpha$ represented an isolated case. We varied $\theta$ between $30^{\circ}$ and $ 90^{\circ}$ in increments of $12^{\circ}$ and again varied $U_c$ from 1.0 to 2.0 in steps of 0.1. For each pair of parameters, 100 independent simulations were conducted with $\alpha=1.5$, and the resulting mean values are shown in the interpolated heat map in figure 4a. As previously mentioned, cluster size is maximized when $\theta=54^{\circ}$ for all values of $U_c$ (note, $\theta=54^{\circ}$ is within the experimentally observed range \cite{Jacobeen_2017}). This is due to a trade-off between local and global packing effects. The number of cells that can pack on a single parent increases with $\theta$--up to $\theta=90^{\circ}$--because the circumference around which daughters are packed is largest at $\theta=90^{\circ}$. However, branches within a cluster interfere with each other less for smaller values of $\theta$; $54^{\circ}$ is the angle where the trade-off between these competing affects is maximized. Additionally, changing $\theta$ (moving it closer to $\theta=54^{\circ}$) is generally a more efficient route to increase cluster size than increasing $U_c$, especially if $\theta$ is far from $\theta=54^{\circ}$. However, since an optimal value of $\theta$ exists (unlike with $\alpha$), when $\theta$ is close to $54^{\circ}$, increasing $U_c$ is more beneficial.

Finally, we investigated the effect of heterogeneity in geometric parameters. Along with providing another geometric parameter to check, monodisperse values of $\alpha$ and $\theta$ are biologically unrealistic, as real snowflake yeast clusters feature polydispersity in both parameters \cite{Jacobeen_2017}. First, a single pair of $\alpha$ and $\theta$ parameters was chosen; we selected $\alpha = 1.5$ because it is in the center of the range of values studied and is within the experimentally observed range, and $\theta = 54^{\circ}$ because it is the optimum value of $\theta$. Variance is introduced in the form of a truncated Gaussian distribution centered on each selected parameter. For every cell added, the value of each parameter is chosen from a self-centered Gaussian distribution; however, if the value selected lies outside the relevant range (1.0 - 2.0 for $\alpha$, $30^\circ$ to 90$^\circ$ for $\theta$), another value is randomly selected. We simulated 100 independent clusters for Gaussians with standard deviations of $0.05$, $0.10$, and $0.20$ of the mean $\theta$ or $\alpha$. 
    
We find that variance in both $\alpha$ and $\theta$ has little effect on cluster size when it is relatively small (standard deviation / mean $\le 0.1$); larger variances, however, ($>0.1$) decrease cluster size (Figure 4b). The inverse relationship between size and large variance is expected for $\theta$; any deviation from the optimal value naturally leads to smaller clusters. However, the relationship between N and $alpha$ is highly linear (figure 2b), meaning that the detriments of smaller aspect ratio cells must outweigh the benefits of longer aspect ratio cells within these disordered clusters. If the standard deviation in alpha decreases from $0.2$ to $0.1$, the resulting increase in cluster size is the same as that caused by an increase in $\alpha$ of $\sim.04$ or and increase in $U_{c}$ of $\sim.26$, again supporting the idea that modifying geometry provides a larger return to the cluster size than modifying bond strength.

\begin{figure}[!thb]
\centering
\includegraphics[width=\linewidth]{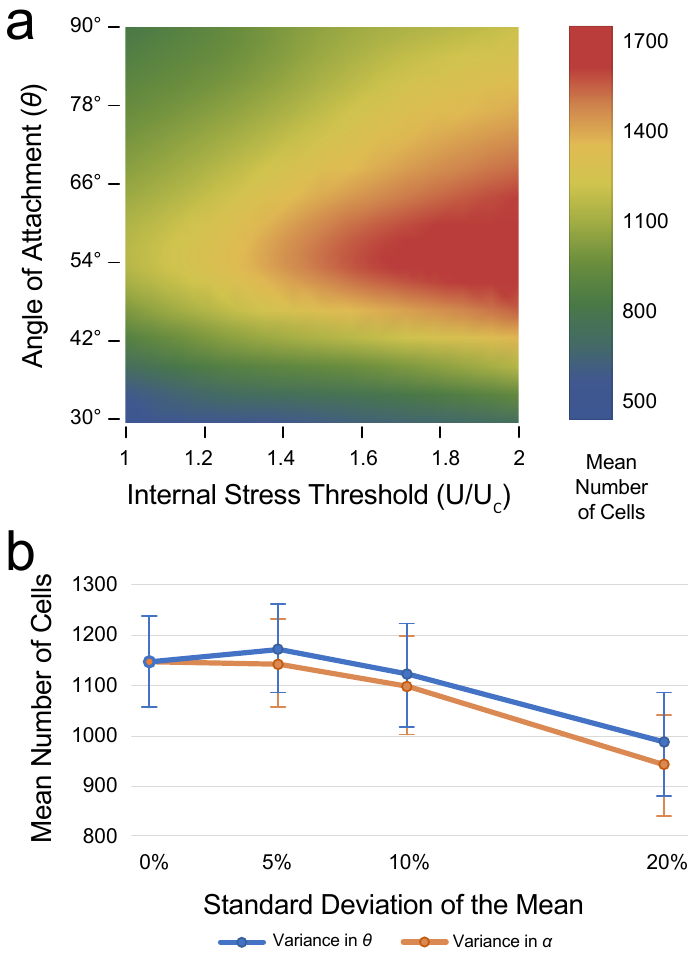}
\caption{(a) Interpolated heat map of the mean number of cells in a cluster as a function of angle of attachment ($\theta$) and deformation energy threshold ($U_{c}$). (b) Effect of variance in the angle of attachment($\theta$) and cellular aspect ratio($\alpha$) on cluster size. The number of cells in a cluster versus the standard deviation of the truncated Gaussian distribution for $\theta$ (blue) and $\alpha$ (orange). Each data point is the average of 100 independent simulations. Error bars indicate standard deviation.}
\end{figure}
\paragraph*{}

\section{Discussion}

Nascent multicellular life was likely under selection for large physical size, as large size provided a number of ecological benefits including protection from predation and toxins \cite{Smukalla2008,BoraasSealeBoxhorn1998,Kessin1996}. However, \textit{how} large physical size could be achieved by newly multicellular organisms has remained poorly understood. Recent work revealed that snowflake yeast evolve increased size via modifications to cellular geometry\cite{Jacobeen_2017}; here, we offer evidence for \textit{why} this route was observed. Geometric modeling reveals that modifying geometry -- via three different parameters -- is a significantly more effective means to achieve larger cluster size than increasing bond strength. Internal stress increases rapidly with reproduction, so investing in bond strength produces diminishing returns. Conversely, modifying cell shape, budding angle, or the variance of these quantities changes how cells pack, slowing the accumulation of internal stress.

Our results highlight the absolute limit of spatial constraints. Two cells cannot overlap, so at high cell density the addition of new cells rapidly increases internal stress. The optimal strategy is not to increase bond strength in the face of vanishing free space, but to pack more efficiently so free space remains available longer. The rapid increase in internal stress with increasing cell number is reminiscent of the jamming transition of athermal grains, for which pressure increases with increasing packing fraction. Previously reported experiments on unicellular yeast demonstrated that reproduction in dense cellular packings can exert pressures on the order of 1 MPa \cite{Hallatschek2016}. Thus, a $\sim 3 \mu m$ diameter bud scar may experience forces on the order of 10 $\mu$N. This is orders of magnitude larger than the $\sim 100$ pN force necessary to break mammalian intercellular bonds \cite{Hosokawa2011,Saumendra_2009} or tear bacteria from a biofilm \cite{Hu_biofilmAFMthesis}. Thus, resisting forces from growth at high cell density would require major innovations on known intercellular adhesion mechanisms.

While snowflake yeast is a lab-evolved model system, it possesses a number of features generally agreed to be common to naturally occurring nascent multicellular organisms. Snowflakes develop clonally, growing through mother-daughter cell adhesion with regular genetic bottlenecks\cite{Ratcliff2012Experimental,Ratcliff2015Origins}. This facilitates multicellular adaptation, as it limits the potential for within-organism genetic conflict and promotes the emergence of novel, heritable multicellular traits \cite{ratcliff2017nascent}. Snowflake yeast readily adapt as multicellular individuals, evolving to be more complex by gaining novel multicellular traits \cite{ratcliff_tempo_2013,Ratcliff2012Experimental,pentz2016apoptosis}. Indeed, complex multicelluarity (\textit{i.e.}, metazoans, land plants, red algae, brown algae and fungi) has only evolved in organisms that develop clonally\cite{brunet2017origin}. Our geometric arguments are easily generalized to other organisms with fixed-geometry morphology. Interestingly, this appears to be the dominant path to complexity: all independent transitions to complex multicellularity, with the exception of animals, grow with rigidly connected cells in a fixed-geometry body plan. Taken together, our results demonstrate that biophysical interactions play a critical role in the evolutionary transition to multicellularity.

\bibliographystyle{apsrev4-1} 
\bibliography{master}

\section{Supplement}

\subsection{Experimental Measurements}

	Experimental measurements used for comparisons with the model in this manuscript were originally reported in \cite{Jacobeen_2017}. Measurements of cellular aspect ratio ($\alpha$) and angle of attachment ($\theta$) were performed on cells from populations of snowflake yeast cells that had been reverted to unicellularity using the lithium acetate / PEG / single-stranded carrier DNA method as described in Ratcliff \textit{et. al.}\cite{Ratcliff2015Origins}. This reversion was accomplished by replacing a single non-functional copy of ace2 (a mutation that arose during experimental evolution) with a functional, ancestral copy. Thus these revertants are thus genetically identical to their snowflake counterparts, with the exception that they are capable of normal mother-daughter cellular separation after mitosis. Revertants were used because they allow for precise measurements of cellular morphology that would have been far more difficult within three-dimensional snowflake clusters. A previous study confirmed that the geometry of revertant cells is not fundamentally different from those within snowflake clusters\cite{Jacobeen_2017}. 
    
    Aspect ratio was measured by imaging several fields of view with a Nikon A1R confocal microscope, and using the particle tracking feature in the image analysis software Fiji. (Statistics: week 1 N = 2128; week 8 N = 1961). To measure the angle of attachment, bud scars (attachment cites) were stained with calcoflour (Fluorescent Brightener 28 from MP Biomedicals, LLC) using the following procedure: a 1:10 dilution of cells from steady state was rinsed and resuspended in deionized water. Then calcoflour was added at a 1:100 dilution from a stock solution of 1 mg/mL calcoflour/water, and this mixture was incubated in the dark at room temperature for at least 10 minutes. Before imaging, the cells were again rinsed and resuspended in deionized water. $\theta$ measurements were obtained from 3D confocal images of individual cells bearing at least 4 bud scars (Statistics: N = 10 for week 1 and week 8).
    
\subsection{Model Validation}

This validation was originally performed in Jacobeen \textit{et. al.}, 2017, and further explanation of these experiments are contained therein\cite{Jacobeen_2017}. To validate the minimal geometric model, we performed simulations utilizing experimentally observed  values of $\theta$ and $\alpha$, and compared several measurements of cluster properties across experiments and simulations. For the validation, $\theta$ was set to 45 degrees (with up to 10\% random variance)--similar to what is observed experimentally\cite{Jacobeen_2017}--and $\alpha$ was randomly seeded with experimentally-obtained cellular aspect ratio distributions obtained from revertant cells from populations of snowflakes that had been isolated after 1, 4, 6, and 8 weeks of evolution. For each distribution, 100 independent clusters were simulated for 12 generations.

\begin{figure}[!thb]
\centering
\includegraphics[width=\linewidth]{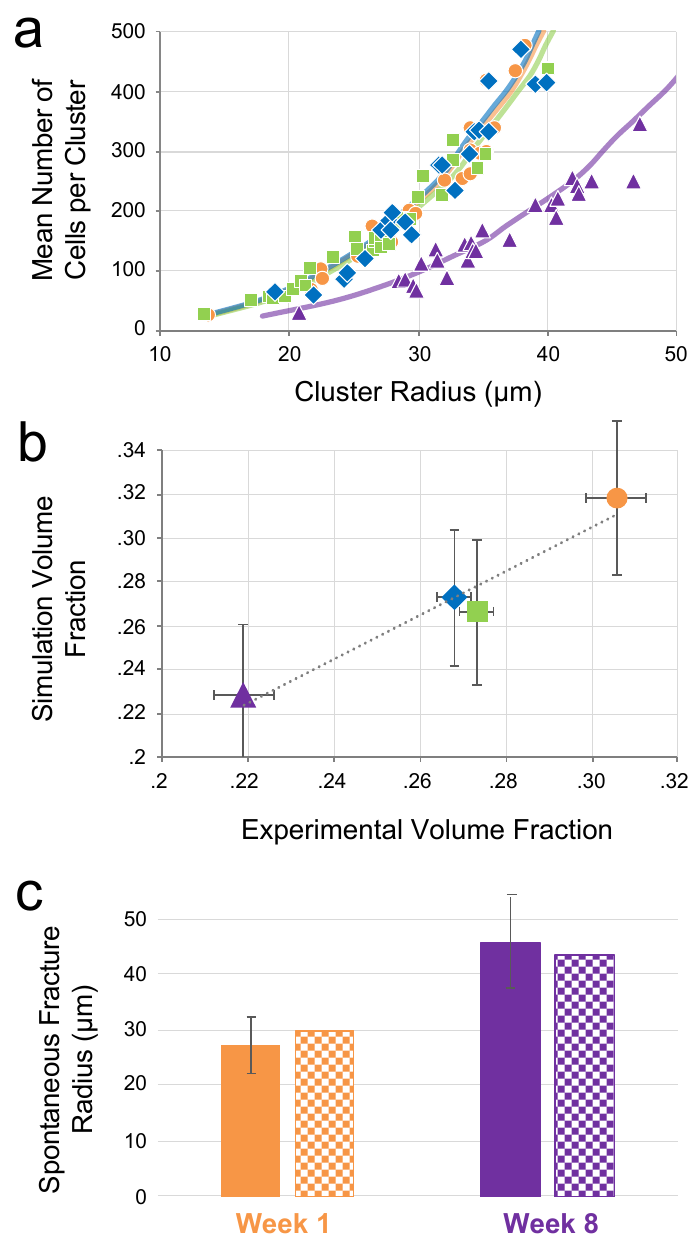}
\caption{(a) Number of cells versus cluster radius for strains isolated after 1 (orange), 4 (green), 6 (blue), and 8 (purple) weeks of daily settling speed selection. Experimental (distinct points) and simulation (continuous lines) are shown, where the simulation data is the mean over 100 clusters. (b) Mean simulation versus mean experimental volume fraction; linear trendline slope $= .998$, $r^2 = .94$. (c) Experimentally observed (solid) and simulation-predicted (checkered) spontaneous fracture sizes for week 1 and week 8 clusters. Error bars represent standard deviation. This data was originally published in Jacobeen \textit{et. al.}, 2017\cite{Jacobeen_2017}}
\label{FIG-PEND}
\end{figure}
\paragraph*{}

	First we assessed the number of cells per cluster as a function of cluster radius. Cluster radius was obtained from a circular approximation of the in-plane area of an intact cluster. Cell counts were then obtained from microscopy images of clusters that had been compressed to a cellular monolayer. Figure 5a shows the remarkable agreement between simulation and experiment in the number of cells as a function of cluster radius for all 4 strains.

	Next, we compared volume fraction, which was obtained by multiplying the number of cells by mean cell volume and dividing by the total volume of the cluster. Again, astounding similarity between experimental and simulated results is observed (Figure 5b); mean volume fraction from simulation is plotted versus that obtained from experiment (trendline slope $= .998$, $r^2 = .94$).

    Finally, we used $U$ to predict fracture size. Mean spontaneous fracture size of week 1 and week 8 strains was obtained from time-lapse microscopy videos of unconstrained cluster growth and fracture\cite{Jacobeen_2017}. By setting $U_{c}$ to the value predicted by the spontaneous fracture size of one strain, the spontaneous fracture size of the other is predicted to well within one standard deviation (Figure 5c).
    
    Collectively, the remarkable agreement between simulation and experiment in number of cells versus radius, volume fraction, and fracture size offers compelling evidence that despite its lack of dynamics, our minimal geometric model accurately describes many of the structural aspects of snowflake yeast clusters. This in turn suggests that physically-imposed geometric constraints play a critical role in determining the structure and fitness of snowflake yeast. 

\end{document}